\documentstyle[12pt,epsfig]{article}
\begin{document}
\begin{center}
{\bf String
Tension and Chern-Simons Fluctuations\\[.1in] in the Vortex Vacuum of d=3 Gauge
Theory}\\John M. Cornwall* and Bryce Yan$^{\dagger}$
\end{center}
\begin{center}{\it Physics Department, University of California at
Los Angeles\\405 S. Hilgard Ave, Los Angeles Ca 90095-1547}\end{center}
\begin{center}{\bf Abstract}
\end{center}

Based on a model of the d=3 SU(2) pure gauge theory vacuum as a
monopole-vortex condensate, we give a quantitative (if model-dependent)
estimate of the relation between the string tension
and a gauge-invariant measure of the Chern-Simons susceptibility, due to
vortex linkages,
in the absence of a Chern-Simons term in the action.  We also give relations
among these quantities and the vacuum energy and gauge-boson mass.  Both the
susceptibility and the string tension come from the same physics:
The topology of linking, twisting, and writhing of closed vortex
strings.  The closed-vortex string is described via a complex scalar
field theory whose action has a precisely-specified functional form,
inferred from previous work giving the exact form of a gauge-theory
effective potential at low momentum.    Applications to high-T
phenomena, including B+L anomalous violation, are mentioned.\\[2in]
UCLA/95/TEP/36 \\
October 1995\\
\footnoterule
\noindent*Electronic address:  cornwall@physics.ucla.edu\\
$\dagger$Electronic address:   yan@physics.ucla.edu
\newpage

\section{Introduction}
In this paper we consider a d=3 parity-conserving (no Chern-Simons
term) pure-gauge (no fermions or scalars) non-Abelian gauge theory,
or just d=3 gauge theory for short.  Our specific considerations will be
for the gauge group SU(2), but the generalization to other gauge groups
is straightforward.

We supply here quantitative details of a scenario proposed earlier
\cite{c95} for d=3 gauge theory, which is strongly-coupled in the
infrared (IR) because the gauge bosons are perturbatively massless.
In this scenario, the gauge-theory vacuum is dominated by an entropy-driven
condensate of closed strings\footnote{The strings we discuss are not the same
as electroweak, or                                                  
Z-, strings\cite{nv}.  Z-strings occur for $T<T_c$ and owe their
existence to a Higgs VEV; they are entropy-disfavored because they
occur in the weak-coupling regime of electroweak theory.} of thickness
$M^{-1}$,
where $M\sim g^{2}_{3}$ is the perturbative gauge-boson mass, proportional
to the d=3 coupling constant $g^{2}_{3}$ (for high=T d=4 gauge theory,
$g_{3}^{2}=g^{2}T$, where $g^{2}$ is the d=4 coupling at temperature T).
Each string has a magnetic flux such that a large Wilson loop in the
fundamental representation:
\begin{equation} W(\Gamma) = P \exp \oint_{\Gamma} d\vec{x}\cdot \vec{A}
\end{equation}
takes on values in the center of the gauge group if the
Wilson loop $\Gamma$ is topologically linked with the closed string,
and is unity otherwise.  (A large Wilson loop is one whose length
scales and distance to the string are $\gg M^{-1}$.)  For SU(2),
the only non-trivial value of W is -1.  The functional average
$\langle W(\Gamma)\rangle$ over all $\vec{A}$ then gives an
area law (described in terms of a string tension), just as for d=4
gauge theory\cite{c79}, where the condensate is formed from closed
two-surfaces.\footnote{This is the mechanism of confinement in non-Abelian 
lattice gauge theory; see, e.g., ref.\cite{t93} and references therein.  In
SU(2) lattice gauge theory, one
introduces monopoles as the site of a junction between a spread-out
(continuum) vortex tube and an infinitesimally-thin $Z_2$ string;
the latter is suppressed in the continuum limit, leaving closed
vortices without monopoles as the means of confinement.}  Note that in (1) we
have absorbed a factor of $i$ (see equation (4) below) as well as the coupling
constant $g_3$ into the definition of the gauge potential.  So $\vec{A}$
has dimensions of mass, while the canonically-defined vector potential,
which is found by dividing $\vec{A}$ by $g_3$, has units of the square root
of mass.

There is another quantity of interest which also depends on linking and
related topological quantities:  The vortex part of the Chern-Simons
susceptibility, which we label $\chi_{CS}$.  We will see that this can be
defined gauge-invariantly and without uninteresting perturbative
contributions.  Begin by defining\cite{j80} the Chern-Simons (CS) number
as usual:
\begin{equation} W_{CS}=\int d^3x\Omega(\vec{x}) \end{equation}
\begin{equation} \Omega = \frac{1}{4\pi^2}\epsilon_{ijk}Tr
(\frac{1}{2}A_i\partial_jA_k-\frac{1}{3}A_iA_jA_k)\end{equation}
where the SU(2) matrices $A_i$ are defined as
\begin{equation} A_i=\frac{1}{2i}A^a_i\sigma_a \end{equation}
Now $\chi_{CS}$ is not gauge-invariant; under a gauge transformation with
winding number N,
\begin{equation} W_{CS}\rightarrow W_{CS}+N \end{equation}
But it is possible to define a gauge-invariant susceptibility as that
part of the zero-momentum correlator
\begin{equation} \int d^3x \langle \Omega(\vec{x})\Omega(0)\rangle
\end{equation}
which is independent of the space volume $V\equiv \int d^3x$ as $V\rightarrow
\infty$ and thus can have no contributions from disconnected parts with
$\langle W_{CS}\rangle=0$, which is a kind of topological neutrality
condition for the string condensate.  The last step in defining $\chi_{CS}$
is simply to restrict the gauge potentials $A_i$ of (3) to those describing
closed strings.  In view of the essentially Abelian nature of these strings
(their holonomy group (1) is Abelian), only the Abelian (first) term of
(3) contributes, and when a string configuration is used to evaluate (6) one
finds\cite{c95,lcs} that it is a sum of Gauss linking numbers for mutually-
linked different strings and twist plus writhe, or self-linking number\cite
{kkn}, for a single string.  This is strictly true if the string centers are
large in the sense previously described for the Wilson loop; at short
distances, there are corrections coming from the gauge-boson mass M which
automatically regulate the usual
divergences associated with self-intersections.  Finally, we can define
\begin{equation} \chi_{CS}=\int d^3x\langle \Omega(\vec{x})                 
\Omega(0)\rangle _{conn}\end{equation}
where only vortex configurations are to be included.  Note that $\chi_{CS}$
scales like $M^3$, or $(g^2T)^3$ in a high-T gauge theory.  Similarly, the
string tension $K_F$ scales like $M^2$, so one can write $\chi_{CS}=
const.(K_F)^{3/2}$.  Our purpose here is to calculate quantitatively this
constant and the ratios $\chi_{CS}/M^3$ and $K_F/M^2$, within the string
condensate model, as well as similar dimensionless relations involving the
vacuum energy and the gauge-boson mass.  If we could calculate any one of
these dimensionful quantities in appropriate units of $g^2T$, all the others
would be determined.

The relation (7) can be written in an interesting way, which has its
counterpart in d=4 gauge theory.  We define the operator $\Theta (\vec{x})$
as the square of the usual (without the coupling constant factor) field
strength:
\begin{equation}\Theta(\vec{x})=\frac{1}{4g_3^2}\Sigma (G^a_{ij})^2(\vec{x}) ;
\end{equation}
note that $\Theta$ is proportional to the trace of the
stress-energy tensor.  We can write (7) as
\begin{equation} \chi_{CS}=\int d^3x \langle \Omega(\vec{x})\Omega(0)\rangle
_{conn}=\frac{\xi}{(4\pi)^4} \langle \Theta \rangle \end{equation}
where the problem is to calculate the dimensionless quantity $\xi$.  One may
check that some such factor as $(4\pi)^4$ is natural, so that $\xi$ is
nominally of $O(1)$.
Equation (8) is the d=3 version of a sum rule conjectured, on various
grounds\cite{cs84}
, for d=4 gauge theory, where $\Omega$ is replaced by the topological charge
density and $\Theta$ is the (anomalous) trace of the d=4 stress-energy tensor.
We can in fact estimate $\xi$ because our work yields an estimate for
$\langle \Theta \rangle$ ins terms of $M^3$.
We have already said that the Wilson-loop area law stems from linking of
condensate strings to a fixed test string, the Wilson loop itself.  There is,
then, a natural relation between the string tension and the CS
susceptibility, which is also expressed in terms of linkages.

Before turning to the string-condensate model, which involves construction
of a complex scalar field theory, let us mention the natural field of
application of these results.  Perhaps the most interesting use of d=3
gauge theory is in electroweak (or other gauge-theory) processes in the
high-T early universe with $T>T_c$, where $T_c$ is the phase-transition
temperature and there is no Higgs mass-generation mechanism.  A finite-T
gauge theory splits into sectors labeled by the Matsubara frequencies
$\omega_N=2\pi NT$; all of the sectors with $N\neq 0$ are well-behaved in
the IR and can be treated more or less perturbatively\cite{zkbn},   if the
coupling is small enough.  Even the N=0 electric sector (gauge potential
$A_0$) generates a perturbative Debye mass and causes no special
difficulties of principle.  But the N=0 magnetic sector, equivalent to
d=3 gauge theory, has IR divergences which must be cured non-perturbatively.

Perturbation theory fails here because no IR-regulating mass can be
generated in perturbation theory; gauge invariance prevents it.  With no
magnetic mass the would-be expansion parameter of high-T theory is
$g^2T/k$, where $k$ is the spatial momentum of the process considered.
This has two consequences: 1) At scales $k$ set by non-zero Matsubara
frequencies, or by the Debye mass $\sim gT$, a genuine perturbation expansion
is possible for small $g$; 2) No such expansion is possible at $k\leq g^2T$,
and
the theory is strongly coupled.
Of course, these IR effects go away below the phase transition where the
gauge bosons get a mass from the Higgs mechanism.

The essence of all non-perturbative effects is the generation of a
magnetic mass $M$ of order $g^2T$\cite{l80,gp82,c82,chk85}.  This mass is,
in principle, to be found by solving some sort of non-linear gauge-invariant
Schwinger-Dyson equation, but analytic progress in this direction has been
slow\cite{chk85,bfhw,eqz93}.  There is a recent lattice determination of
this mass\cite{hkr95}, but the technique used is not quite gauge-invariant.
We will not attempt to determine $M$, but given its existence we will analyze
further non-perturbative effects which can be expressed in terms of it
(see the discussion below equation (6)).

These effects include high-T sphalerons\cite{mketc,p93} and vortex strings
which can either close or terminate on magnetic monopoles; we need only
consider closed strings.  These are like Nielsen-Olesen strings except
that there is no Higgs effect and no symmetry breaking.  Because d=3 gauge
theory is strongly coupled, these strings form an entropy-driven condensate.
The existence of a condensate follows directly from certain exact results
\cite{c94} for d=3 gauge theory, which include among others $\langle \Theta
\rangle >0, \epsilon_{vac}=-(1/3)\langle \Theta \rangle <0$\cite{s93}, where
$\Theta$ is the squared field strength (see (8)) and $\int d^3x\epsilon_{vac}$
is the vacuum energy (or equivalently the N=0 contribution to $\beta F$
in the thermal case, with F the Helmholtz free energy)\footnote{Shaposhnikov,
in ref.\cite{s93}, also attempts to estimate the free energy by semiclassical
means; we will compare his value to ours in the conclusions.}  Of course,
the negative sign of $F$ reflects entropy domination.

The fluctuations in CS number measured by the susceptibility (7) play
an important role in determining the rate of anomalous B+L violation\cite{krs,
am87,ckn93} from
high-T electroweak processes in the early universe.  However, it is not
possible to determine this B+L rate/unit volume $\Gamma_{BL}$ solely from
the essentially static fluctuations of $W_{CS}$ in the d=3 gauge-theory
sector; one really needs to understand fluctuations in the rate of change
$\dot{W}_{CS}$ of CS number, which are directly related to fluctuations
of topological charge (and B+L, by the anomaly equations).  To calculate
$\Gamma_{BL}$ requires much more than d=3 gauge theory, and we will not
attempt such a calculation here.  The general wisdom\cite{krs,am87} is that
for $T>T_c, \Gamma_{BL}$ scales like $(\alpha_WT)^4$, that is, like $M^4$,
where M is the magnetic mass.  This is just dimensional analysis, if $M$ is
the only relevant scale; the hard part is to calculate the overall
coefficient.  Usually one argues\cite{krs,am87} that there is no sphaleron
barrier for $T>T_c$ so the coefficient is O(1), but this is not really
clear.  Given a magnetic mass M, there should (Cornwall,
ref.\cite{mketc};Philipsen\cite{p93})
be a sphaleron of mass $M_{Sp}\sim M/\alpha_W \sim T$, and the thermal
activation
factor $\exp (-M_{Sp}/T)$ is the exponent of a negative pure number, which
is not necessarily small in absolute value.  We
will not consider sphalerons further here, but it will certainly be important
to estimate their contributions to $\chi_{CS}$, etc.\footnote{Actually,
sphalerons are closely related to the vortices
we use (J. M. Cornwall and G. Tiktopoulos, Phys. Lett. B181, 353 (1986);
M. Hindmarsh and T. W. B. Kibble, Phys. Rev. Lett. 55, 2398 (1985)).
A sphaleron interpolates between regions of oppositely-directed magnetic
flux on a vortex, and can be pulled apart into two monopoles of the type
that can terminate a thick vortex string\cite{t93}.}

In our vortex model, $\Gamma_{BL}$ depends on the rate of topological
reconnection\cite{dletc} of strings in the condensate.  This dynamical
process involves electric gauge fields, which lie outside the d=3 gauge-theory
sector we study here, and we will not discuss it in detail. However,
we will make a few remarks on how the d=3 strings are promote to closed
two-surfaces in d=4 and on how an intersection-number\cite{mn86} topological
invariant of these surfaces is related to the usual topological charge.

Now let us return to pure d=3 gauge theory and our method of approach to
its non-perturbative effects.  It has been known for decades\cite{est}
that a condensate of strings can be described by a (possibly complex)
scalar field theory, with a wrong-sign mass term representing entropy
effects and an interaction term like $\lambda \phi^4$ representing the
repulsion, or increase in free energy, of two strings trying to occupy
the same region of space.  The underlying connection between a string
condensate and a scalar field theory is that a field-theoretic propagator
can be represented as a sum over paths, or strings, and in particular
the field-theory vacuum functional becomes a partition function of
closed strings.

Such a connection between a d=3 string condensate and a field theory would
not have been useful in the past, because there would have been more parameters
for the scalar field theory than for the underlying gauge theory.  However,
we will exploit here the exactly-known\cite{c94} form of the zero-momentum
effective action for the operator $\Theta$ of equation (8).  Given this
effective action, we show that the corresponding classical action for the
scalar field theory is uniquely\footnote{We will see later why a complex rather
than a real field is                                                       
chosen.}of the form
\begin{equation} S=\int d^3x [|\vec{\nabla} \psi |^2 + \frac{\lambda |\psi
|^4}{4} ] \end{equation}
where the coupling constant $\lambda$ depends on $M$.  From this classical
action we will derive loop terms, automatically consistent with the
gauge-theory effective action, which are essential to describe the condensate
entropy.  These loop terms simply modify the classical action of (10) to
an effective action
\begin{equation} \Gamma = \int d^3x [ |\vec{\nabla} \psi |^2 +
\frac{\lambda |\psi^4 |}{4} - \frac{\lambda v |\psi |^3}{3} ] \end{equation}
where $v$ is the expectation value of $|\psi |$, also expressible in terms of
$M$.
Like all effective actions, (11) should be treated classically.  Although we
can and do derive the form (11) from loop corrections to the classical action
(10), the functional form of (11) is completely dictated by the gauge-theory
effective action.  Note that (11) has a phase with $\langle |\psi | \rangle
= v$ and short-range fluctuations, as required to describe the gauge-string
condensate.

Given this complex scalar field theory, how does one describe the
topological effects involved in the string tension and the CS susceptibility?
In both cases, one uses auxiliary Abelian gauge potentials, but in different
ways.  For the string tension the auxiliary gauge potential is itself a
fixed vortex, described in terms of the Wilson loop contour; this fixed
potential is coupled to the scalar field theory in the usual way (which
requires a complex field).  At large distances--all that is relevant for
the string tension--this Wilson loop potential becomes a pure gauge of the
type $\vec{\nabla} \Lambda$, but $\Lambda$ is singular and $\exp i\Lambda$
is not single-valued.  (This is because we seek the fundamental-representation
string tension; for the adjoint representation, $\exp i\Lambda$ is
single-valued and the string tension vanishes.)  The multi-valuedness is in
conflict with the necessary single-valuedness of the complex scalar field,
or equivalently the underlying d=3 gauge potentials, and when one minimizes
the scalar-field effective action in the presence of the auxiliary gauge
potential one finds that this conflict is resolved by a surface of zeroes
of the scalar field, coinciding with the minimal spanning surface of the
Wilson loop.  As a result a string tension $\sim M^2$ is generated; the
numerical coefficient is an elliptic integral.

To calculate the CS susceptibility, we use a technique given long ago by
Edwards\cite{e68} to study topological entanglement of polymers.  A new
Abelian gauge potential $\vec{V}(\vec{x})$ is introduced, with a standard
coupling to the scalar field and a self-action which is a pure CS term
(except for very important short-distance modifications, coming from the
gauge mass $M$, which make everything finite).  Functional integration over
$\vec{V}$ yields a power series in the V-field coupling constant, whose
coefficients are related to expectation values of Gauss linking integrals.
In particular, we can read off the CS susceptibility from this series.
To construct the series to the requisite (fourth) order we need to
calculate a large number of three-loop vacuum graphs, all of which are
finite and scale like $M^3$.  We have calculated the sum of graphs mostly
numerically, after reducing the graphs to Feynman-parameter form.

\section{Scalar Field Theory Description of the Vortex Condensate}

The usual discussion\cite{est} of mapping a string condensate onto a
field theory begins with the observation that the number of configurations
of a closed string on a lattice, of length $l$ and step length $L$, is
roughly (ignoring unimportant effects) given by the path integral

\begin{equation} \oint (d\vec{x})\exp -\int_0^l (\frac{\dot{x}^2}{2}-\ln 2d )
dl' /L \end{equation}
up to an irrelevant normalization.  Here $d$ is the number of dimensions
in which the string lives, and the entropy factor $(2d)^{l/L}$ is roughly the
number of ways the string can turn at its $L/l$ vertices, without
consideration of self-avoidance effects. If the strings are oriented, as they
are for us, one multiplies this
result by $\int dl/l\exp -Ul$ and exponentiates to get the string partition
function.  Here $U$ is the internal energy/unit length of the string.
(The $l$ in the denominator of the integral adjusts for overcounting the
point on the string arbitrarily chosen as a starting point.)  The final
result is recognized as the functional integral of a free
two-component real
(or one complex) scalar field, where the logarithm of the propagator of the
field has been expressed as a proper-time integral.  The field has a mass term
which may be normal or tachyonic, depending on whether the internal energy
dominates the entropy or otherwise.  A string condensate will form only if
the entropy dominates, as it does for d=3 gauge theory.

Next one must add self-avoidance effects, which amounts to adding interaction
terms for the scalar field inside the functional integral.  It is, of course,
natural to have a four-field interaction to represent the simplest kind of
intersection of two strings, but the most general renormalizable field theory
in d=3 has other types of interactions, ranging up to sixth order terms.
The question is what terms and what coupling constants are needed.

We will address this problem in the reverse order of the remarks above,
where the possible tachyonic mass term (signaling a string condensate)
is conceptually introduced before self-avoidance effects.  Based on earlier
exact results for the gauge-theory effective action\cite{c94}, we will
construct a classical scalar action containing a fourth-order interaction
but no mass terms at all.  We will then show that one-loop effects contain
a negative contribution to the effective action, equivalent to a tachyonic
mass term.  The resulting effective action has a minimum corresponding to
a string condensate.

The idea behind the gauge-theory results is very simple:  There is only
one coupling constant, $g_3^2$, and it has dimensions of mass.  The action
of the gauge theory can be written as:
\begin{equation} S_g=\int d^3x \Theta =\frac{1}{4g_3^2}\int d^3x
\Sigma (G^a_{ij})^2 \end{equation}
(where $\Theta$ is the squared field strength of equation (8)).
The partition function Z, which is the functional integral of $\exp -S_g$,
gives $\exp -\int d^3x \epsilon
_{vac}$.  By naive dimensionality, $\epsilon_{vac}\sim g_3^6$, and repeated
differentiation of Z and of $\epsilon_{vac}$
with respect to $g_3^2$ yields an infinite set of sum rules for
the zero-momentum vacuum matrix elements of $\Theta$.  The functional
differentiation of Z acts only on the explicit $g_3^{-2}$ in the action (13),
and the specific form of the sum rules, or effective action which generates
them, depend only on the specific powers of $g_3$ in Z and $\epsilon_{vac}$.
The effective action can
be found by introducing a constant source $J$ into the functional integral,
defining a vacuum functional {\bf W} as follows:
\begin{equation} Z= \exp -{\bf W}(J)=\int (dA) \exp [-S_g (1-J)] \end{equation}
                                            Because the only dependence on
$g_3^2$ is an overall inverse factor in the action, adding
the source is the same as changing $g_3^2$ to $g_3^2 (1-J)^{-1}$.  Then
\begin{equation} {\bf W}(J)=\epsilon_{vac} (1-J)^{-3} \end{equation}.
By the usual Legendre transform one finds the effective action for the
zero-momentum matrix elements of $\Theta$:
\begin{equation} \Gamma (\Theta )=\int d^3x (\Theta - \frac{4}{3}
\Theta^{3/4}\langle \Theta \rangle^{1/4}) \end{equation}
One sees that $\Gamma$ has a minimum at $\Theta = \langle \Theta \rangle$
of value $-\int d^3x (1/3)\langle \Theta \rangle=\int d^3x \epsilon_{vac}$.

What are the consequences for the scalar theory?  A litle thought shows that
this theory must have only one coupling constant $\lambda$, and this coupling
can be chosen to have dimensions of mass.  Then in order to reproduce the form
of the effective action as in (16), $\epsilon_{vac}$ must
scale like $\lambda^3$, corresponding to the scaling of the gauge action as
$g_3^6$, and and it must be possible to rescale $\psi$ so that the only
appearance of $\lambda$ in the action is an overall factor of $\lambda^{-1}$.
So the scalar action (10)
\begin{equation} S=\int d^3x [|\vec{\nabla} \psi |^2 + \frac{\lambda |\psi
|^4}{4} ] \end{equation}
is acceptable.  Of course, other interaction terms could be included,
such as $\lambda^{3/2}|\psi^3|$, but they will not do.  The reason is that
with these other interaction terms  one cannot rescale $\psi$ in such
a way that the action has an overall factor of $\lambda^{-1}$.  But such a
rescaling is essential to the derivation of the gauge-theory effective
action.

The upshot is that only the classical action (17) can yield loop corrections
consistent with the gauge theory effective action (16).  There is one subtlety
of the action (17): It generates quadratic and sextic terms from loops, which
appear to be divergent and require bare terms in the action to accommodate
the needed renormalizations.  However, demanding consistency with the
gauge-theory effective action fixes these renormalizations.  A simple way to do
this is to introduce a free energy ${\bf W}_s(J)$ for the scalar theory,
analogous to ${\bf W}(J)$ for the gauge theory introduced above.  Here J is a
constant source for the scalar action S.  Then we will require
\begin{equation} {\bf W}_s(J)=(1-J)^{-3}{\bf W}_s(0), {\bf W}_s(0)=\epsilon
_{vac},
\end{equation}
just as for the gauge theory as shown in equation (15).

Next we turn to the calculation of one-loop corrections to the classical
action (17), which express the entropy of strings.  Rather than calculate
${\bf W}(J)$ as outlined above, we will directly calculate the effective
action (essentially the free energy) using the CJT\cite{cjt} formalism.
Following the one-loop calculation we will estimate two-loop contributions,
which are appreciable, and presumably suggest the accuracy of  the calculations
(perhaps $\pm 30\%$).  Please note that the CJT approach does not deal with
bare loops, that is, perturbative effects, but instead with dressed loops
constructed from the full propagator.  To start, write
\begin{eqnarray} Z & = & \exp (-\beta F) = \int (D\psi D\bar{\psi})
\exp (-S)\\ & = & \exp \{ tr \ln \frac{G}{G_0}-tr(GG_{\psi}^{-1}-1)
-\int\frac{\lambda}{4}|\psi^4 | + 2PI graphs \}  \end{eqnarray}
Here G is the full propagator of the theory, $G_{\psi}$ is the classical
propagator in the field $\psi$, and 2PI graphs are connected graphs with more
than
one loop that are two-particle irreducible. The notation tr indicates an
integral over all space, or a sum over all momenta.  The physical free energy
is found
by searching for the extrema of $\beta F$ in the functional variable G as
well as in the field $\psi$.  To this end, we introduce a variational
parameter $m$ in G, writing in momentum space
\begin{equation} G^{-1}=k^2+m^2. \end{equation}                           
We are only interested in small-momentum phenomena, so $\psi$ can be treated
as a constant, and one readily finds the classical propagator
\begin{equation} G_{\psi}^{-1}=k^2+\lambda |\psi |^2. \end{equation}
All one-loop graphs, with G as the propagator, are found by dropping the
2PI graphs in (19).  The next step is to insert G and $G_{\psi}$ into the
first two terms of (19) and do the integrals.  Divergences appear, which can
be removed in either of two equivalent ways:  Dimensional regularization, or
by the substitution
\begin{equation} \int d^3k F(k^2) \rightarrow -2\int d^3k(1+k^2 \frac{\partial}
{\partial k^2})F \end{equation}
It is easily checked that this substitution leaves convergent integrals
unchanged, and eliminates divergences.

Before removing the divergences, we should comment on their physical meaning
in the case of interest, finite-$T$ gauge theory.
The underlying d=3 gauge theory has no divergences, except those which can
be interpreted as coming from the other sectors which go to make up the
full finite-$T$ theory.  For example, any divergence in the coupling constant
would have to come from an underlying d=4 divergence.  In general, ultraviolet
divergences are associated with scales such as $T$, rather than $g^2T$ which
is appropriate to the d=3 gauge theory itself.  We are justified in
regulating the scalar theory in a way which reproduces known results of the
d=3 gauge theory, such as the effective action form (16).

It is a remarkable property of three dimensions that integrals whose integrands
depend on $m^2$ yield odd powers of $m$.  In fact, these odd powers must be
understood as absolute values, in view of the symmetry $m\rightarrow -m$ of
the integrand.  Ultimately, these powers of $m$ will yield odd powers of
$|\psi |$, as in (11).  Moreover, the one-loop integrals after regularization
yield negative contributions to the effective potential.  For further
discussion and more references, see ref.\cite{btt93}.

Now do the integrals in (19), following the above prescriptions, to find
\begin{equation} \beta F=\int d^3x[\frac{\lambda |\psi |^4}{4}-\frac{|m|^3}
{6\pi}-\frac{|m|}{4\pi}(\lambda |\psi |^2-|m|^2)]. \end{equation}
Upon varying $m$ one finds\footnote{This value of $m$ means that $G=G_{\psi}$,
as required by
the variational equation for $G$ which comes from equation (20) when the
2PI graphs are omitted.}$m^2=\lambda |\psi|^2$, and substituting yields
\begin{equation} \beta F=\int d^3x [\frac{\lambda |\psi |^4}{4}
-\frac{\lambda^{3/2} |\psi |^3}{6\pi}]. \end{equation}
Except for the gradient terms, which will be considered later, this is of
the form (11) and consonant with the gauge-theory effective action (16),
provided
that one identifies
\begin{equation} \Theta=\frac{\lambda |\psi |^4}{4}, \langle \Theta \rangle
=\frac{\lambda^3}{4(2\pi)^4}. \end{equation}
Further minimization on $\psi$ yields:
\begin{equation} \langle |\psi | \rangle \equiv v = [\frac{m}
{2\pi}]^{1/2}, m=\frac{\lambda}{2\pi}, \epsilon_{vac}=V^{-1}\beta F_{min}=
-\frac{m^3}{24\pi}.\end{equation}
Here $V$ is the volume of all space.
It is important to note that the variational parameter $m$ is not the physical
scalar mass value,
which is found from the second derivative of the effective action at the
minimum.  We require that the scalar mass, or correlation length, must be
the same as for the d=3 gauge theory, and we will use the same notation $M$
for both of these quantities.  It is easy to find from (25) that
\begin{equation} M^2=\frac{1}{2}\lambda v^2 = \frac{1}{2}m^2.\end{equation}

Next we ask what the effect of two-loop terms is.  There is only one
2PI two-loop graph, the double bubble, and its value, to be added to
the integrand of the one-loop value (24), is $+\lambda m^2/32\pi^2$.
As before, vary to find $m$, with the result:
\begin{equation} m=-\frac{\lambda}{8\pi}+ [\lambda \psi^2 + (\frac
{\lambda}{8\pi})^2]^{1/2}. \end{equation}.
When this result is inserted in the two-loop effective action, new
terms involving $\psi$ appear which are not of the desired form (11).  This
is no problem of principle; once one goes beyond one-loop graphs, one really
cannot recover the desired form (11) without considering loops of all orders.
The reason is that a generic multi-loop graph corresponds to an expansion
in powers of $\lambda /\psi^2$, as one readily checks, and such terms are
already found in the expansion of m in (29).  A simple way to include
the needed parts of other graphs to correct the explicit two-loop result
has already been given in (18):  One calculates the true vacuum energy
$\bf W_s(0)$
with zero source J at two loops, and then writes the vacuum functional
in the presence of the source as in (18) by multiplying by $(1-J)^{-3}$.
After some uninteresting algebra, one finds the following two-loop
parameters:
\begin{equation} \epsilon_{vac}= -\frac{m^3}{24\pi}, v=(\frac{m}{2\pi})^{1/2},
m=M. \end{equation}
Note that, in spite of a very different algebraic structure, the expression
of $\beta F_{min}$ and $v$ in terms of $m$ are the same for the one- and
two-loop
effective actions; the only difference is the relation between the physical
mass $M$ and the variational parameter $m$, which differ in the two cases by
a factor of $\surd 2$.  This can be taken as a rough measure of the (not very
high) accuracy of the first few terms of the CJT loop expansions, although the
error encountered in any specific quantity may be more or less than this.
Perhaps this is an acceptable discrepancy in analytic calculations of a
strongly-
coupled gauge theory, where there is no obvious small parameter in the
kind of dressed-loop expansion we are using.

The point here is not to show that the effective scalar action must have the
form (11), which is guaranteed by the underlying gauge theory, but to
estimate the parameters $\lambda, v$ which occur in it in terms of the
physical mass $M$.  This we have done in equations (27, 28, 30).  The next step
is to calculate the string tension $K_F$ in terms of the same parameters.

\newpage
\section{The String Tension}

To calculate the string tension via the expectation value of the Wilson
loop
\begin{equation} \langle W(\Gamma)\rangle = \langle Tr P\exp \oint_{\Gamma}
d\vec{x}\cdot \vec{A} \rangle \end{equation}
we write for $\vec{A}$ a representative vortex configuration and sum over
the collective coordinates of the vortices.  This will be recognized as the
coupling of a string condensate to a fixed Abelian vortex gauge potential
generated
by the Wilson loop contour itself.  The usual rules\cite{est} for converting
the string expectation value to the scalar field theory generate a standard
minimal gauge coupling of the Wilson-loop vortex to the scalar field.
Classical minimization of the resulting effective action leads to
differential equations whose solution gives rise to the string tension.

As has been discussed elsewhere\cite{c79,c95} the gauge-field vortex is a
classical solution to an effective gauge action\footnote{This effective
action summarizes some of the quantum effects which generate the mass, but
to the extent that it generates new short-distance (scales $\ll M^{-1}$)
singularities it is not suitable for use at such scales. To the extent that the
mass is important in shielding short-distance singularities, as we find in the
present work, there is no reason to doubt the correctness of the effective
action.  Consistency requires not
only that the mass $M$ vanish at short distances, but also that there are
other group-singlet scalar excitations, or glueballs, which do not need
discussion in this paper.} which includes a gauge-invariant mass
term\cite{c74,c82}.  This action is:

\begin{equation} S_v = S_g - \int d^3x \frac{M^2}{g_3^2} Tr(V^{-1}\vec{D}V)^2
\end{equation}
where $S_g$ is the usual massless gauge action (see (13)), $V$ is an
auxiliary SU(2) matrix, and $\vec{D}$ is the covariant derivative
$\vec{\nabla}+\vec{A}$.  The gauge transformation laws
\begin{equation} \vec{A} \rightarrow U\vec{A}U^{-1} + U\vec{\nabla}U^{-1}
\end{equation}
\begin{equation} V \rightarrow VU \end{equation}
show that the mass term in (32) is locally gauge-invariant (it is just a
gauged non-linear sigma model).  The V-field can be functionally integrated
out at the classical level, just by solving the V-field equations of motion.
These turn out to be the same as the vanishing of the covariant derivative of
the $\vec{A}$
equations of motion.

Up to a local gauge transformation, the vortex solution can be written:
\begin{equation} \vec{A}(\vec{x}) = 2\pi Q \vec{\nabla}\times \oint
d\vec{z} [\Delta_M (\vec{x}-\vec{z}) - \Delta_0 (\vec{x}-\vec{z})]
\end{equation}
where Q is an SU(2) generator such as $\sigma_3/2i$ with the property that
$\exp 2\pi Q$ is in the center of SU(2).  Here $\Delta_M$ is the scalar
propagator for mass $M$:
\begin{equation} \Delta_M(\vec{x}) = \frac{1}{(2\pi )^3}\int d^3k \frac
{e^{i\vec{k}\cdot \vec{x}}}{k^2 + M^2} \end{equation}                      
and $\Delta_0$ is the corresponding massless propagator.  The loop integral
is over a closed string which describes the center of the vortex, whose
field strength extends a distance $\sim M^{-1}$ from this center.  However,
the potential itself has a long-range pure-gauge part associated with
$\Delta_0$; this part comes from the $V$-field in the effective action
(32).  The long-range pure-gauge part is, as we will see\cite{c79},
responsible for confinement and the string tension.  The pure-gauge term
has its role at short distances too, where the leading singularities at
$\vec{x}\sim \vec{z}$ cancel between the two propagators in (35).  This
will be crucial in deriving finite results for the CS susceptibility, but
is not important for the string tension; the only singularities cured by
a mass term are in perimeter-law pieces of the Wilson loop.

The Wilson-loop expectation now becomes
\begin{equation} \langle W\rangle =\langle \langle exp \pm i\pi \oint
_{\Gamma}d\vec{x}\cdot \vec{\nabla}\times \oint_s d\vec{z} [\Delta_M-
\Delta_0]\rangle \rangle \end{equation}
 where the double brackets indicate an expectation value over the string
partition function.  The contour $\Gamma$ is, as before, the Wilson-loop
contour, and $s$ labels a set of string contours.

At this point we can appreciate the connection between confinement and
linking of the Wilson loop with a closed string.  The concept of string
tension is appropriate for a large Wilson loop, with scales $\gg M^{-1}$;
this means that we can drop the $\Delta_M$ term in (37).  The remaining
term in the exponent immediately becomes $\pm i\pi L$, where $L$ is the
Gauss linking integral:
\begin{equation} L= \oint_{\Gamma}\oint_s d\vec{x}\times
d\vec{z}\cdot \frac{(\vec{x}-\vec{z})}{4\pi |\vec{x}-\vec{z}|^3}
\end{equation}
The quantity $L$ measures (with sign) the number of times the contours
$s$ and $\Gamma$ are linked.  Clearly, an even (odd) number of linkages
contribute a term $\exp(i\pi L)= +1(-1)$
to the Wilson loop.  The Wilson loop is large, so $L$ is a sum of a large
number $N$ of random positive or negative integers, and $L$ has a Poisson
distribution.
Moreover, $N$ is proportional to the (minimal) area A of the Wilson loop,
since all unlinked strings give no contribution to the string tension:
$N=\rho A$, where $\rho$ is a two-dimensional density of strings in the
vacuum.  It is then an elementary excercise to calculate the expectation
of $\exp(i\pi L)$ in the Poisson distribution (which we approximate by a
Gaussian), with an area law as the
result:
\begin{equation} \langle \langle \exp(i\pi L) \rangle \rangle
=\exp(-\pi^2N/2) =\exp(-\pi^2 \rho A/2)\equiv \exp(-K_F A) \end{equation}
where $K_F$ is the string tension.
We see that the string tension measures in some sense the fluctuations
of linkages of the string condensate with the fixed string of the Wilson
loop, so it is natural to expect a close connection between the string
tension and another measure of linkage fluctuation given by the
CS susceptibility.

Note, by the way, that had we desired $\langle W\rangle$ in the adjoint
representation, $\pi$ in (39) would be replaced by $2\pi$ and there would
be no string tension.

It is difficult to estimate the string density $\rho$ in a straightforward
elementary way, so instead we turn to the description of the string
condensate by the scalar field theory.  Observe that the Wilson-loop
expectation value (37) has a standard gauge action $S_W$ coupling the strings
to the
fixed Abelian gauge potential $\vec{W}$:
\begin{equation}S_W=i \oint d\vec{z}\cdot \vec{W} \end{equation}
\begin{equation} \vec{W}(\vec{z}) = \pi \vec{\nabla}\times \oint
_{\Gamma} d\vec{y} [\Delta_M(\vec{z}-\vec{y})-\Delta_0(\vec{z}-\vec{y})]
\end{equation}
This is of the form of the original gauge vortex in (35), with the
Wilson-loop contour $\Gamma$ as the center of the vortex.
Such a coupling, expressed in scalar-field language through the proper-time
formalism, amounts to writing the scalar effective action given in (11)
with a gauge-covariant derivative:
\begin{equation} S_W= \int d^3x [ |\vec{D}\psi |^2+\frac{\lambda |\psi^4 |}{4}
- \frac{\lambda v |\psi |^3}{3} ] \end{equation}
where the covariant derivative is
\begin{equation} \vec{D}=\vec{\nabla}-i\vec{W}. \end{equation}
So the Wilson-loop value is
\begin{equation} \langle W\rangle = Z^{-1}\int (D\psi D\bar{\psi})
\exp(-S_W) \end{equation}
where $Z$ is the scalar-field partition function in the absence of the
Wilson-loop gauge potential.  At the classical level, $Z$ is just
\begin{equation} Z=\int (D\psi D\bar{\psi})\exp(+\int d^3x
\frac{\lambda v^4}{12}), \end{equation}
that is, a functional integral over the action evaluated at the $\vec{W}
=0$ saddle point $S=-\int d^3x \frac{\lambda v^4}{12}$.
Then we can write the logarithm of the Wilson loop, or $K_F A$, as
\begin{equation}K_FA= -\ln \langle W\rangle =  \int d^3x [ |\vec{D}\psi
|^2+\frac{\lambda |\psi^4 |}{4} - \frac{\lambda v |\psi |^3}{3}+\frac{\lambda
v^4}{12}] \end{equation}
The idea is to find a classical solution (i.e., minimum of the action)
and to read off the string tension from the above integral.

As mentioned above, the Wilson loop is large and effects associated with
the $\Delta_M$ term in $\vec{W}$ of (41) do not contribute to the area-law
part of the action, only to curing short-range divergences in perimeter-law
contributions.  Let us see what happens when we drop this term.  The
remaining $\Delta_0$ term of $\vec{W}$ is pure gauge, as can be shown
directly with Stokes' theorem; it corresponds to the $V-$term in (32).  A
pure-gauge term can normally be compensated by choosing the phase of the field
to which the gauge potential is coupled, so
it is tempting to say that, if
\begin{equation}   \vec{W}=\vec{\nabla}\Lambda \end{equation}
then a solution to the classical equations is just
\begin{equation} \psi = ve^{i\Lambda} \end{equation}                       
for which the action integral (46) vanishes.  This argument is globally quite
correct
for the string tension in the adjoint representation, for which $\Lambda$ is
twice as big as in the fundamental representation, but not for the
fundamental representation.  The reason is that in the fundamental
representation $\Lambda$ of (47), which is multi-valued, is such that
$\exp(i\Lambda)$ is two-valued when either (but not both)
of the points $x=\pm a$ are encircled in the $xy-$plane.  Multi-valuedness of
$\psi$ is not allowed; like the gauge potentials themselves, this field
must be single-valued.

We now show that there is another solution for the fundamental representation,
which has a finite string tension.  The first step is to specify the
Wilson loop, which is taken to be two infinite straight lines parallel to the
z-axis
and passing through the points $x=\pm a, y=0$.  We require $Ma\gg 1$ so that
the Wilson loop is large, justifying the dropping of $\Delta_M$ in $\vec{W}$.
One readily calculates
from (41) the gauge function $\Lambda$ of (47):
\begin{equation} \Lambda(\vec{x})=\frac{1}{2}[\phi_{a}-\phi_{-a}]
\end{equation}          
where
\begin{equation} \phi_{\pm a}=\arctan \frac{y}{x\mp a} \end{equation}
It is, of course, the 1/2 in (49) that makes $e^{i\Lambda}$ multi-valued, and
which would be missing in the adjoint representation.
We can make $e^{i\Lambda}$ single-valued in a cut plane, by giving
$\phi_{\pm a}$ each a cut along the positive $x$-axis, starting at
$x=\pm a$.  Then the gauge function $e^{i\Lambda}$ has no cut (is
single-valued) for $|x|>a$,
but its phase has a jump of $\pi$ across the portion of the x-axis joining the
two sides of the
Wilson loop.  We can explicitly exhibit the structure of $e^{i\Lambda}$
on this cut, and it is convenient to do so at $x=0$:
\begin{equation} e^{i\Lambda}=\epsilon (y)\frac{y+ia}{\surd y^2+a^2}
\end{equation}
which changes sign from i to -i upon crossing the $x$-axis, and approaches
unity at $y=\pm \infty$.  This change of sign holds not only at $x=0$ but
all along the $x$-axis.

Now separate $\psi$ into a real part and a phase:
\begin{equation} \psi=Re^{iQ} \end{equation}
Here R is real, but can be positive or negative.  We can choose $Q=\Lambda$
outside a circle in the $xy$ plane of radius $r\gg a$, where there is no cut,
but we cannot do so for $|x|<a$  because of the jump discontinuity. However, it
is possible to
find another solution for $|x|<a$ to the equation
\begin{equation} \vec{\nabla}Q=\vec{\nabla}\Lambda \end{equation}
which is valid  for $|x|<a$ except on the cut at $y=0$.  This solution, unlike
$e^{i\Lambda}$ of (51), has no jump.  It is sufficient for the gradients to
agree in order to compensate for $\vec{W}$ in the covariant derivative.  At
$x=0$ we can again exhibit this solution explicitly:
\begin{equation} e^{iQ}=\frac{y+ia}{\surd y^2+a^2} \end{equation}
There is no cut, and $e^{iQ}$ approaches unity at $y=+\infty$.  But for
$y=-\infty$, this phase factor approaches -1.

Consider the behavior of $\psi$ as $r\rightarrow \infty$.  In this domain
we can take
\begin{equation} \psi=ve^{i\Lambda} \end{equation}
as the classical solution, up to terms exponentially small in $Mr$.  This is
just what we have already done for $|x|>a$, but not for $|x|<a$.  In the latter
case the phase factor $e^{iQ}$ has opposite signs in the upper and lower
half planes, which is inconsistent with choosing the real factor $R=v$.
Clearly $\psi$ cannot change sign for sufficiently large $r$ without leading
to infinite action (46).  The simple solution is to allow $R$ to change sign
too, cancelling the sign change in $e^{iQ}$.  This sign change is
brought about by requiring $\psi$ to vanish along the cut $|x|<a$, y=0,
which we do by requiring $R$ to be odd in $y$ for $|x|<a$.  This is consistent
with the field equations.  Given that (53) holds, the gauge potential drops out
of the equations of motion.  In these equations, the $x-$derivatives are only
important in a region of $O(M^{-1})$ around the Wilson loop, so we keep only
the $y-$derivatives.  Then the equation of motion is:
\begin{equation} -R''+\frac{\lambda}{2}R^2(R-\epsilon (R)v)=0
\end{equation}
Here the primes indicate $y$-derivatives and
the $\epsilon$ factor comes from differentiating $|\psi|^3$.  We choose $R$
to have the same sign as $y$, and find the solution for $y>0$, x=0 which
vanishes at $y=0$:
\begin{equation}y= \int_0^R dx [2(\frac{\lambda}{4}x^4-\frac{\lambda v}{3}
x^3+\frac{\lambda}{12}v^4)]^{1/2} \end{equation}
Expansion of $R$ in (56) around y=0, plus $R(-y)=-R(y)$, shows that $R$ has
continuous derivatives through third order at the origin.

So far we have constructed some useful pieces of a full solution to the
classical equations of motion.  To find an exact solution would require
some smoothing and matching along the lines $x=\pm a$, but we will not
do that here because it contributes nothing to the string tension.
For $|x|>a$ the approximate solution is $\psi=ve^{i\Lambda}$, while for
$|x|<a$ the solution is $Re^{iQ}$ as shown in (54, 56).  These must be
patched together in a neigborhood of size $\sim M^{-1}$ around $|x|=a, y=0$,
where the two solutions differ, and this would require keeping the
$\Delta_M$ term in $\vec{W}$.

The string tension-area product is given in the action integral (46).
The area factor arises as a product of $\int dz$ times the integral over
$x$ from $-a$ to $a$, and the string tension $K_F$ is the remaining factor, an
integral over $y$.  Some standard manipulations which convert the integral
over $y$ to an integral over $R$, plus some rescaling, put the string tension
in the form
\begin{equation} K_F=4Mv^2\int_0^1 du [\frac{1}{6}(3u^4-4u^3+1)]^{1/2}
\end{equation}
This elliptic integral yields
\begin{equation} K_F=1.17Mv^2  \end{equation}                              
Using the one- and two-loop results of equations (27, 28, 30) we can
convert this into $K_F=const.M^2$; numerical results will be discussed in
the concluding section, after we find the CS susceptibility in the next
section.
\newpage

\section{The Chern-Simons Susceptibility}

First, we observe that the CS number $W_{CS}$ (see (2,3)) associated with
strings or vortices can also
be expressed as a modified Gauss linking number, just as the string tension
can.
The modifications are due to the finite mass $M$ and come from the
$\Delta_M$ term of the vortex (35); they vanish for linkages of large
well-separated strings, but are vital to regulate string intersections,
including
self-intersections.  We insert the vortex form (35) into the CS number (2,3)
to find for the CS number coming from two distinct vortices (labeled 1 and 2):
\begin{equation} W_{CS}(1,2)=-\frac{1}{8\pi^2}\int d^3x Tr[\vec{A(1)}\cdot
(\vec{\nabla}\times \vec{A(2)})+(1\leftrightarrow 2)] \end{equation}
Insert the vortex expression (35) to come to
\begin{equation} W_{CS}(1,2)=\frac{\pm 1}{2}\oint_1 d\vec{z_1}
\times\oint_2 d\vec{z_2} \cdot \frac{\vec{R}}{4\pi R^3} F(MR) \end{equation}
In this equation
\begin{equation} \vec{R}=\vec{z_1}-\vec{z_2} \end{equation}
and the function $F$ comes from the massive and massless propagators in
the vortex expression (35):
\begin{equation} F(R)=\frac{1}{2}\int_0^{MR} du u^2 e^{-u} \end{equation}
For large well-separated strings $MR\gg 1$ and $F\rightarrow 1$, so
\begin{equation} W_{CS}=\pm \frac{1}{2}L(1,2) \end{equation}
           where $L(1,2)$ is the Gauss linking number of the two strings.  But
as
$MR\rightarrow 0, F\sim(MR)^3$ which completely removes the singularity
in the Gauss integral when $\vec{R}=0$.

The CS number of (64) is fractional, but one should expect for the
gauge group SU(N) to encounter CS numbers which are multiples of 1/N;
this is equivalent to periodicity $2\pi N$ in the dependence on the
vacuum angle.  Such dependence is not in contradiction to the requirement
that the CS number be an integer, which is a global requirement;
the linking of any two particular strings is local.

Self-linking of a single string also contributes to the CS number an
amount $(1/2)W_{CS}(1,1)$ (see (61)).  For a large string with no
self-intersections and minimum chord length $\gg M^{-1}$, one again recovers
a Gauss linking integral which, as is well-known\cite{kkn}, expresses the
self-linkage as the sum of two integers, twist and writhe (which separately
have no invariant topological meaning).  For a string with self-intersections
or near misses the
function $F(MR)$ regulates possible divergences, somewhat in the spirit of
the usual ribbon framing procedure.  However, the strings are fat, and
the self-linking CS number can take any value.  Again, this does not
interfere with the global requirement that the total CS number be an integer.

We wish to calculate the fluctuation in linkages of the string condensate.
The first step, following Edwards\cite{e68}, is to write the sum over strings
of (61) in
propagator form:
\begin{equation} W_{CS}=\frac{1}{4}\int d^3x \int d^3y J_i(\vec{x})
\Delta_{ij}(\vec{x}-\vec{y}) J_j(\vec{y}) \end{equation}              
where
\begin{equation} \Delta_{ij}(\vec{x})=-i\int d^3k e^{i\vec{k}\cdot \vec{x}}
\epsilon_{ijl}k_l\frac{M^4}{k^2(k^2+M^2)^2} \end{equation}           
and the current is the usual one:
\begin{equation} \vec{J}(\vec{x})=\sum_{strings} \oint d\vec{z} \delta
(\vec{x}-\vec{z}) \end{equation}                                    
Now introduce an Abelian gauge field $\vec{V}(\vec{x})$ for which $\Delta$
is the propagator, and a functional integral over $\vec{V}$ which
generates expectation values of powers of $W_{CS}$:
\begin{equation} Z_V(\zeta)=\int (DV)
\exp\{i\int [\frac{1}{2}V_i\Delta^{-1}_{ij}V_j+\zeta \vec{J}\cdot
\vec{V}]\}\equiv \exp(-\omega
(\zeta)). \end{equation}                                                
(Strictly speaking, $\Delta$ has no inverse, and a gauge-fixing term should
be added to (68) in order that the inverse exist.  This term contributes
nothing and will be omitted from the explicit discussion.)
The action for $\vec{V}$ is of CS type, but with a propagator modified at
short distances according to (66); at long distances, this propagator is
just the usual CS one.  The coupling constant $\zeta$ is introduced to
allow construction of matrix elements of $W_{CS}$ by differentiation,
after doing the functional integral over $\vec{V}$.  This integral is, up
to an irrelevant normalization,
\begin{equation} Z_{\zeta}=\langle \langle exp(\frac{-i\zeta^2}{2}
\int \int J\cdot \Delta \cdot J)
\rangle \rangle \end{equation}                                         
where, as in (37), the double brackets indicate a string expectation value.
Then from (65, 69) the connected CS susceptibility $\xi_{CS}$ (see (7)) is
\begin{equation} \frac{1}{4V}(\frac{\partial}{\partial \zeta^2})^2 \omega 
(\zeta)|_{\zeta =0} \end{equation}
where V is the volume of all space.

As before, we calculate $\omega(\zeta)$ in the scalar-field form, which
amounts to changing the scalar action (17) to
\begin{equation} S_{\zeta}=\int d^3x [|\vec{D_{\zeta}} \psi |^2 + \frac{\lambda
|\psi
|^4}{4} ] \end{equation}                                                  
where the new covariant derivative is
\begin{equation} \vec{D_{\zeta}}=\vec{\nabla}+i\zeta \vec{V} \end{equation}
Then $Z_{\zeta}$ is found from a functional integral over $\psi$ and $\vec{V}$.
While the effect of the $\vec{V}$ field can be calculated perturbatively,
we already know that non-perturbative effects of the $\psi$ field are crucial.
Once again we use the CJT formalism to express these effects, including
graphs with two $\vec{V}$ lines (thus of $O(\zeta^4)$) in the 2PI sum (see
(20).
The calculation will be done
on-shell (that is, $\psi \rightarrow v$, the physical expectation value).
As before we express the free energy as the sum of one-particle terms and
2PI graphs:
\begin{equation} -\omega(\zeta)=tr\{\ln\frac{G}{G_0}-\frac{G}{G_0}+1
+\frac{1}{2}\ln\frac{\Delta}{\Delta_0} -\frac{1}{2}(\Delta \Delta_0^{-1}-1)\}
+\Sigma \end{equation}
and use a simple massive form for the dressed propagator $G$, as in equation
(21), with variational parameter $m$:
\begin{equation} G(k)=(k^2+m^2)^{-1}.\end{equation}                
 The term $\Sigma$ represents the graphs shown in Fig.1, where the solid lines
represent the propagator $G$. This propagator differs from our previous
solution by powers of $\zeta$, and is to be determined from its variational
equation.  The solution is a non-perturbative one as far as $\psi$ goes
(which we have worked out at one- and two-loop level earlier; see Section 2),
modified by $\vec{V}$ corrections.

     The result is the set of graphs shown in Fig. 2, where
the solid lines now represent a massive propagator with the {\it physical}
mass $M$.  A couple of
comments are in order:  1) The weights given in the figures include the
factor of 1/2 required for closed loops consisting of one or two $\vec{V}$
lines; 2) All the {\it bare} graphs of $O(\zeta^2)$ vanish, because they
contain a single $\epsilon$-symbol.  This means that the expectation value of
the CS number is zero, as
one expects for a parity-conserving theory.  However, the corresponding
$O(\zeta^4)$
{\it dressed} graphs do not vanish, since they have two $\epsilon$ symbols.

It remains to evaluate these graphs, which have conventional Feynman rules
except for the $\vec{V}$ propagator, which is given by (66).  Note that this
falls off like $k^{-6}$ in momentum space, so rapidly that all the graphs
we need to calculate have no divergences from $\vec{V}$ lines.  (We regulate
any $\psi$ divergences as before, using (23).)  The evaluation is
straightforward but lengthy, and cannot be done completely analytically.
Our approach is to introduce Feynman parameters and to do the momentum-space
integrals analytically, then to do the remaining Feyman-parameter integrals
numerically.  There are integrable singularities in the Feynman-parameter
integrals, but these cause no difficulties.  In certain cases some of the
Feynman-parameter integrals can be done analytically, but the generic case
is a seven-fold integral (there are usually eight parameters, constrained by
a delta-function, because of the nonstandard form of the $\vec{V}$ propagator
in (66)).  Another approach to this propagator is to write it as:
\begin{equation} \frac{1}{k^2}(\frac{M^2}{k^2+M^2})^2=
\int_0^{M^2} d\sigma \frac{1}{(\sigma +M^2)^2} - \frac{M^2}{(k^2+M^2)^2}
\end{equation}                                                             
This is not necessarily an easier way to do things, but it affords a check
on our results since it provides an alternative Feynman-parameter form
for numerical integration.  Both ways give the same answer.

The final result for the CS susceptibility, expressed in terms of $M$, is
(using (70))
\begin{equation} \chi_{CS}= 0.010\frac{M^3}{(4\pi)^4} \end{equation}       
In the concluding section below, we will evaluate this and other results
in terms of the string tension and the vacuum energy.
\newpage
\section{Numerical Results and Conclusions}
First we discuss the numerical results of our calculations, and then
briefly point out how reconnection of strings leads to topological charge
change.

For purposes of comparison with lattice and other calculations, it is
useful to express all quantities in terms of the string tension $K_F$, whose
value is fairly accurately known from lattice work\cite{t92,ikps}.  This value
is
\begin{equation} K_F=0.11-0.13 (g^2T)^2 \end{equation}
(here and in what follows we always write $g^2T$ for $g_3^2$, in view of
applications to high-$T$ gauge theory).  Below we show values of
$\epsilon_{vac}$, $M$, and $\chi_{CS}$ in appropriate units of $K_F$ and
$g^2T$, at one and two loops, using a nominal value $K_F^{1/2}=0.36g^2T$.
These results are based on equations (27, 28, 30, 59, 76).
We use superscripts $(1), (2)$ to indicate one-and two-loop values.

\begin{eqnarray} \epsilon_{vac}^{(1)}&=&-0.28K_F^{3/2}=-0.013(g^2T)^3\\    
\epsilon_{vac}^{(2)}&=&-0.17K_F^{3/2}=-8\times 10^{-3}(g^2T)^3\\
M^{(1)}&=&1.95K_F^{1/2}=0.70g^2T\\M^{(2)}&=&2.32K_F^{1/2}=0.83g^2T\\
\chi_{CS}^{(1)}&=&0.074\frac{K_F^{3/2}}{(4\pi)^4}=\frac{3.5\times 10^{-3}}
{4\pi}(\alpha_WT)^3\\
\chi_{CS}^{(2)}&=&0.12\frac{K_F^{3/2}}{(4\pi)^4}=\frac{5.6\times 10^{-3}}
{4\pi}(\alpha_WT)^3 \end{eqnarray}

One can get a rough idea of the errors involved in truncating the loop
expansion from these numbers; they depend on the dimensionality of the
quantity involved.  For $M$, with dimensions of mass, the error is about
30\%, with errors in other quantities growing as the dimension grows.

There are only a few other calculations one might compare to these numbers.
Concerning the non-perturbative vacuum energy,
refs \cite{ikps} and \cite{kkrs} give values for $-\epsilon_{vac}$
in the range $0.016-0.027(g^2T)^3$, while Shaposhnikov\cite{s93} gives
$0.033(g^2T)^3$.  The first two references cited above are lattice works for
the full electroweak theory at finite $T$ (near the transition temperature)
and include Higgs and $N\neq 0$ modes.  However, these should not contribute
substantially to a non-perturbative quantity likes $\epsilon_{vac}$.
Ref. \cite{s93} is based on a dilute-monopole gas approximation, and contains
some factors arbitrarily taken to be unity.  One sees here a wide spread in
these other calculated values, which are not too far from the results we
calculate.

For the gluon mass $M$, a recent lattice determination\cite{hkr95} gives
$M=0.46g^2T$, somewhat smaller than we give.  This is not a pole
mass determination, which would be gauge-invariant, but we do not expect
there to be a serious gauge dependence in this number.  Refs \cite{bfhw,eqz93}
give a small value $M=g^2T/3\pi = 0.11g^2T$ from a continuum one-loop gap
equation, but again this value is not gauge-invariant and one has no reason
to believe that a one-loop gap equation with no vertex corrections is at all
reliable.  Ref \cite{chk85} gives a lower limit on $M$ of about $0.58g^2T$,
which is consistent with our values.  Ref\cite{chk85} uses a non-linear {\it
gauge-invariant}\cite{c82} one-dressed-loop
gap equation with vertex corrections and the seagull graph included.  The
reason
\cite{c82} that only a lower limit can be given is that at two-loop order
there is a logarithmic divergence in the seagull graph even in perturbation
theory, and this must (because there is no mass counterterm) be cancelled
by other two-loop contributions which nobody has studied yet.  In other words,
without imposing gauge invariance on a two-loop or higher gap equation,
one has no control over perturbative effects which lead to (ultimately
cancelling) ultraviolet divergences.  So the authors of ref. \cite{chk85}
showed that their non-linear equation had no solution at all unless $M$ were
a certain minimum value, no matter what happened to the seagull graph.

There are no other computations known to us of $\chi_{CS}$ with which our
result
can be compared.  Various authors have estimated a related quantity, which is
$\Gamma_{BL}$, the rate at which $B+L$ is violated, or equivalently (up
to a factor of $N_F$, the number of flavors)  the rate at which the CS number
diffuses.  Given the static mean-square fluctuation $X_{CS}$, one can
estimate $\Gamma_{BL}$ by multiplying by a rate.  If we take this rate to be
$M$ and ignore factors which we hope are $O(1)$, this leads to
\begin{equation} \Gamma_{BL}/N_F \approx M\chi_{CS}=7-8\times 10^{-3}
(\alpha_WT)^4 \end{equation}                                               
Philipsen\cite{p93} has done a calculation of $\Gamma_{BL}$ in high-T
electroweak theory, based on the proposed existence (Cornwall, ref
\cite{mketc}) of sphalerons above the phase transition temperature, where
the Higgs VEV vanishes. His result maximizes at about $0.01(\alpha_WT)^4$ at
$M=0.1g^2T$, and falls off rapidly on either side.  The dependence of
Philipsen's result on $x\equiv M_{Sp}/T$ is complicated, but somewhat similar
to the usual semiclasasical sphaleron rate below the transition temperature
\cite{am87}:
\begin{equation} \Gamma_{BL} \sim (\alpha_WT)^4x^7e^{-x} \end{equation}
If, as given in refs \cite{c77,c95}, $M_{Sp}\approx 5.3M/\alpha_W$, one finds a
maximum rate at $M\approx 0.1g^2T$ ($x=7$), a mass value proposed before\cite
{bfhw,eqz93}.  But if the values of $M$ in equations (80, 81) are used,
the quantity $x$ is very large, about 45-50, and Philipsen's rate would be very
small.  Given such sensitivity of sphaleron rates to $M_{Sp}$, it is simply not
clear
yet whether high-$T$ sphalerons are or are not important, but our numbers
taken at face value would not leave much room for sphalerons to dominate
B+L decay above the phase transition temperature.

We close by showing how the conventional d=4 topological charge is related
to a change in vortex linkage.  For d=4 the form of the vortex corresponding
to the d=3 form (35) has long been known (see, e.g., ref \cite{c79}).
It is:
\begin{equation} A_{\mu}(x)=2\pi Q\epsilon_{\mu \nu \alpha \beta}
\partial_{\nu}\oint d\sigma d\tau\frac{1}{2}(\dot{z_{\alpha}}z\prime_{\beta}
-(\alpha \leftrightarrow \beta))[\Delta_M(z-x)-\Delta_0(z-x)] \end{equation}
                       where the $\Delta$s are d=4 propagators, $Q$ is a group
generator as in
(35), and $z(\sigma ,\tau)$ describes a closed two-surface.  This potential is
a solution to the d=4 action analogous to the d=3 action in (32), and in the
static limit ($z=(\vec{z}(\sigma ),\tau )$) reduces to (35).  When two strings
cross each other in such a way as to change their linkage (Gauss integral),
the topological charge $q$ expresses this change as a d=4 integral giving the
intersections of the two closed surfaces.  We give the result only for two
large surfaces, where the $\Delta_M$ term can be neglected:
\begin{equation} q\equiv -\frac{1}{16\pi^2}\int d^4x\mbox{ } trG_{\mu
\nu}\tilde{G}
_{\mu \nu}=\frac{1}{2}\int d\sigma_1 d\sigma_2 d\tau_1 d\tau_2
\epsilon_{\mu \nu \alpha \beta} \dot{z}_{1\mu}z\prime_{1\nu}\dot{z}
_{2\alpha}z\prime_{2\beta}\delta(z_1-z_2)
 \end{equation}
Here $z_1(\sigma_1,\tau_1)$, $z_2(\sigma_2,\tau_2)$ are the equations of the
two surfaces.  This integral clearly measures the (generically pointlike)
intersections of the two surfaces.  A similar integral can be written for
self-intersections (see also ref. \cite{mn86}), and the finite-$M$ corrections
worked out.  The topological charge for
SU(2) is half-integral, like the CS number, but it is a local contribution
to the global charge, which must be integral, and is no obstruction to an
integral global charge.\footnote{For a construction of localized instantons
of half-integral charge, tied together by a sphaleron world line, see
Cornwall and Tiktopoulos as cited in footnote 3.}  The reader can easily
construct kinematic configurations for the surfaces which correspond to
the time evolution of a change of linkage, and see how the topological
charge is generated thereby.  Of course, such time evolution necessarily
involves electric fields and goes beyond the d=3 gauge theory we have
considered here.

There still remains much to be done on d=3 gauge theory before we can
have any confidence in attacking such difficult dynamical problems as
B+L washout at high $T$.  The next significant step will be do construct
a Schwinger-Dyson equation for the gauge-boson mass $M$ which, like the
work reported in Cornwall {\it et al.}\cite{chk85}, is gauge-invariant and
includes
vertex parts consistent with the Ward identities, but which goes to at
least two dressed loops in order to deal with incipient perturbative
two-loop divergences in the mass.  Then one must face up to the problem
inherent in all analytic treatments of strongly-interacting gauge
theories, with no obvious expansion parameter:  How accurate are the results?
The only self-contained approach is to keep more 2PI graphs in the CJT
effective potential.  Aside from that, we can only compare with lattice
computations.  It would be valuable to have a lattice calculation of the
CS susceptibility for this purpose.

\newpage
\centerline{\bf Acknowledgments}
JMC was supported in part by the NSF under Grant PHY-9218990,
and BY received partial support from the UCLA Committee on Research.

\newpage
\begin{center}{\bf Figure Captions}\end{center}
Fig. 1.  Dressed-propagator graphs for the 2PI graphs $\Sigma$ (defined in the
text).  The wavy lines
are $\vec{V}$ propagators, and the solid lines are dressed $\psi$
propagators, including dressing with $\vec{V}$ lines as shown in Fig. 2.

\vspace{12pt}
\noindent Fig. 2.  Graphs of Fig.1 expanded to show $\vec{V}$ lines occurring
in the
propagator $G$ of that figure.  The solid line now means the physical
$\psi$ propagator in the absence of $\vec{V}$.
\newpage
\epsfig{file=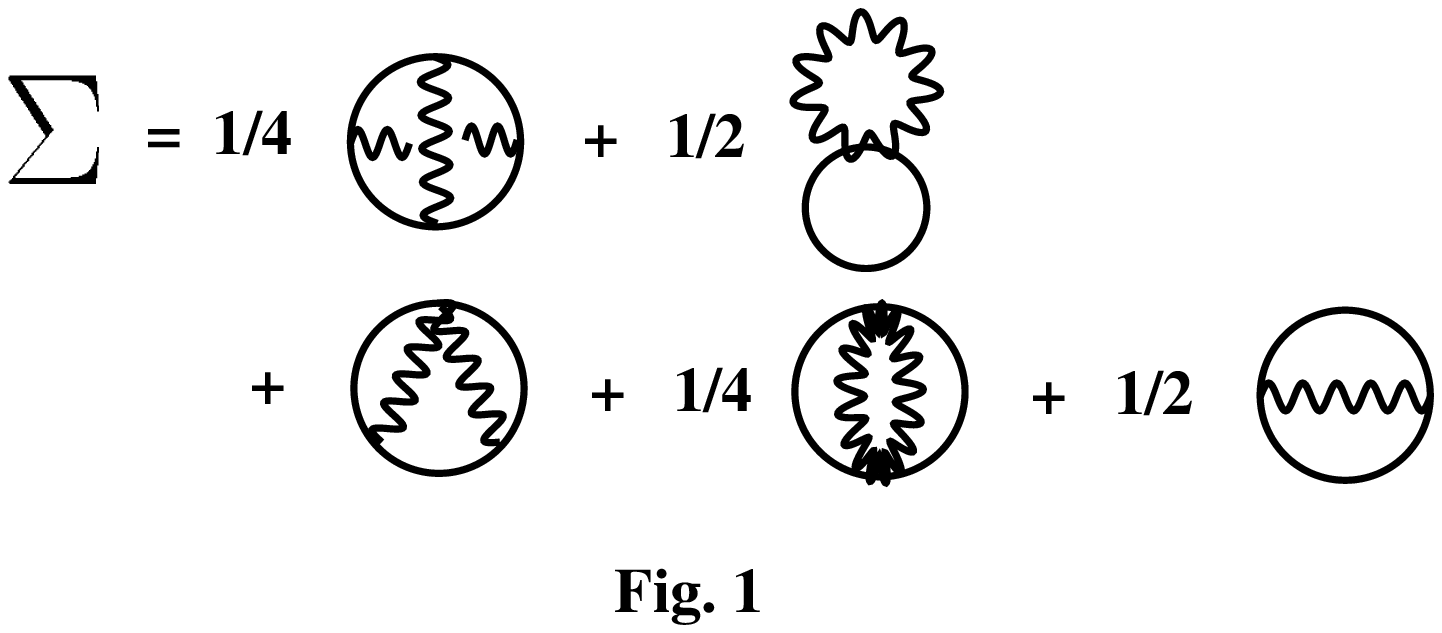,clip=}
\vspace{.5in}
\epsfig{file=yan2.eps,clip=}

\end{document}